\documentclass[11pt]{article}
\usepackage{moriond,epsfig}

\def\Journal#1#2#3#4{{#1} {\bf #2}, #3 (#4)}

\def\PRL{\em Phys. Rev. Lett.}
\def\PRB{{\em Phys. Rev.} B}

\def\be{\begin{equation}}
\def\ee{\end{equation}}
\def\bea{\begin{eqnarray}}
\def\eea{\end{eqnarray}}

\begin{document}
\title{Observation of $h/e$ conductance oscillations in disordered metallic 
$\mathcal{T}_{3}$ network}

\author{\underline{F. Schopfer}$^1$, F. Mallet$^1$, C. Naud$^{2}$, G. 
Faini$^3$, D. Mailly$^3$, L. Saminadayar$^{1,4}$ and C. B\"{a}uerle$^1$}

\address{$^1$Centre de Recherches sur les Tr\`es Basses
Temp\'eratures, B.P. 166 X, 38042 Grenoble Cedex 09, France}
\address{$^2$Laboratoire d'\'{E}tudes des Propri\'{e}t\'{e}s \'{E}lectroniques
des Solides, B.P. 166 X, 38042 Grenoble Cedex 09, France}
\address{$^3$Laboratoire de Photonique et Nanostructures, route de
Nozay, 91460 Marcoussis, France}
\address{$^4$Universit\'{e} Joseph Fourier, B.P. 53, 38041 Grenoble
Cedex 09, France}

\maketitle

\abstracts{We report on magnetotransport measurements performed on a
large metallic two-dimensional $\mathcal{T}_{3}$ network. 
Superimposed on the conventional Altshuler-Aronov-Spivak (AAS)
oscillations of period $h/2e$, we observe clear $h/e$ oscillations in
magnetic fields up to $8\,T$. Different interpretations of this
phenomenon are proposed.}

\section{Introduction}

Quantum interferences resulting from the phase coherence of the
electronic wave functions lie at the heart of mesoscopic physics. In
a mesoscopic ring pierced by a magnetic flux, such interferences lead
to the well known Aharonov-Bohm (AB) conductance oscillations: the
conductance oscillates with magnetic flux\cite{Buttiker} with a period
$\phi_{0}=h/e$, where $h$ is the Planck constant and $e$ the charge of
the electron.

In a line of rings, ensemble averaging leads to a strong suppression
of these $h/e$ oscillations due to the random phase of the
oscillations in each ring. On the other hand, interferences due to
time reversed trajectories do not average to zero\cite{Altshuler} and
give rise to the so-called Altshuler-Aronov-Spivak (AAS) oscillations
of period $h/2e$. This has been beautifully demonstrated
experimentally in series of experiments on silver
rings\cite{Washburn}.

The same phenomenon should take place in a two dimensional array of
rings. However, only the limit of very large number of rings has been
explored experimentally\cite{Pannetier}: in this case, one clearly
observes $h/2e$ oscillations around zero field, but no $h/e$
oscillations.

Renewal of interest in interference phenomena in such networks has
recently arisen from the theoretical study of networks of a specific
geometry called $\mathcal{T}_{3}$ networks. In such a geometry, it
has been suggested that interference effects should lead to a
localisation of the electrons into \emph{Aharonov-Bohm cages} when the
magnetic flux is exactly half a flux quantum per unit
cell\cite{Vidal1}. Such a localization leads to a strong enhancement
of the $h/e$ oscillations even in large networks\cite{Vidal2}. This
has been demonstrated experimentally on networks made from high
mobility heterostructures\cite{Naud}: clear $h/e$ oscillations have
been observed even at high magnetic field. However, such an
experiment only tests the low disorder, small number of channels
limit.

In this article, we report on magnetotransport measurements on
metallic $\mathcal{T}_{3}$ networks. Superimposed on the usual $h/2e$
conductance oscillations around zero magnetic field, we observe clear
$h/e$ conductance oscillations. These oscillations are of strikingly
similar amplitude as the $h/2e$ AAS oscillations around zero magnetic
field and persist at high magnetic field of $8\,T$ with no significant
decrease in amplitude.

\begin{figure}[hbt]
\rule{5cm}{0.2mm}\hfill\rule{5cm}{0.2mm}
\vskip 0.5cm
\begin{center}
\psfig{figure=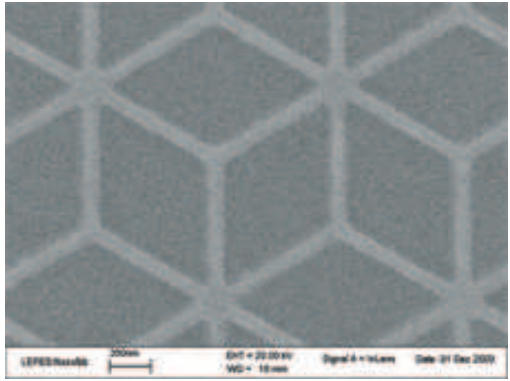,height=5cm}
\end{center}
\rule{5cm}{0.2mm}\hfill\rule{5cm}{0.2mm}
\caption{\textsc{sem} picture of the $\mathcal{T}_{3}$ lattice.}
\label{fig:sample}
\end{figure}

\section{Measurements}\label{subsec:prod}

Samples were fabricated on silicon substrate using electron beam
lithography on polymethyl-methacrylate resist. The metal is deposited
using an electron gun evaporator and lift-off technique. A $1\,nm$
thin titanium layer is evaporated prior to the gold evaporation in
order to improve adhesion to the substrate. For the gold evaporation,
we used a source of $99.999\,\%$ purity.

The sample consists of a $\mathcal{T}_{3}$ lattice containing $4500$
unitary cells. Wires have a length $a=690\,nm$, width $w=80\,nm$ and
thickness $t=30\,nm$ (see figure \ref{fig:sample}), corresponding to a
flux quantum per unit cell of $\phi_{0}=100\,G$. The resistance of
the network is $24\,\Omega$ at $4.2\,K$. The elastic mean free path
is evaluated to be $l_{e}=28\,nm$ and the diffusion coefficient,
extracted from $D=1/3v_{F}l_{e}$, is $D=1.3\cdot
10^{-2}\,m^{2}s^{-1}$. As expected, the wires are quasi one
dimensional with respect to both the phase coherence length $l_\phi$
and the thermal length $L_T = \sqrt(\hbar D/k_{B}T)$. Resistance
measurements have been carried out in a dilution refrigerator using a
standard $ac$ resistance bridge technique.

\begin{figure}[htb]
\rule{5cm}{0.2mm}\hfill\rule{5cm}{0.2mm}
\vskip 0.5cm
\begin{center}
\psfig{figure=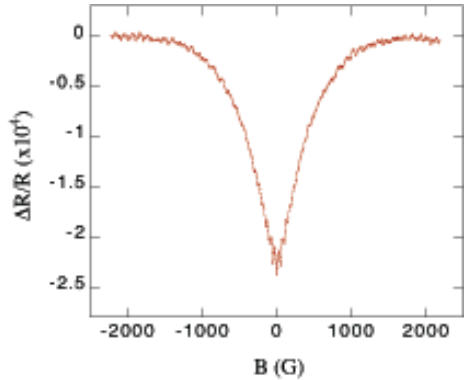,height=5cm}
\end{center}
\rule{5cm}{0.2mm}\hfill\rule{5cm}{0.2mm}
\caption{Magnetoresistance around zero field of a metallic 
$\mathcal{T}_{3}$ lattice at a temperature of $400\,mK$. The 
absolute resistance of the sample is $24\,\Omega$.} 
\label{fig:low-field}
\end{figure}


\section{Results and discussion}

We first concentrate on low field measurements. Figure
\ref{fig:low-field} shows the relative magnetoresistance of the
$\mathcal{T}_{3}$ lattice around zero magnetic field at a temperature
of $400\,mK$. From the envelope of the weak localization signal, we
evaluate a phase coherence length of $l_{\phi}\approx 2\,\mu m$. 
Superimposed on this large magnetoresistance, we see clear
oscillations of period $50\,G$ around zero field, corresponding to
$\phi_{0}/2$ per unit cell. At higher magnetic field, small
oscillations appear.

\begin{figure}
\rule{5cm}{0.2mm}\hfill\rule{5cm}{0.2mm}
\vskip 0.5cm
\begin{center}
\psfig{figure=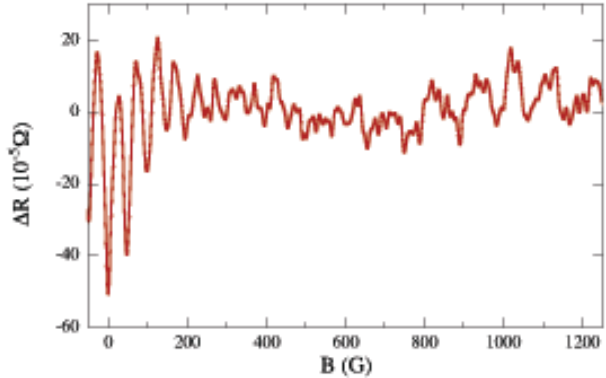,height=5cm}
\end{center}
\rule{5cm}{0.2mm}\hfill\rule{5cm}{0.2mm}
\caption{Magnetoresistance around zero field of a metallic 
$\mathcal{T}_{3}$ lattice at $400\,mK$, after subtraction of the weak 
localisation contribution.} 
\label{fig:low-field-fit}
\end{figure}

To highlight these oscillations at high field, we have subtracted from
the magnetoresistance the weak localisation correction. Figure
\ref{fig:low-field-fit} displays the obtained magnetoresistance. 
Again, at magnetic field below $150\, G$, one clearly observes AAS
oscillations of amplitude $\Delta R=6\cdot 10^{-4}\Omega$,
corresponding to $2.6\cdot 10^{-2}\,e^{2}/h$ in terms of dimensionless
conductance. As expected, these AAS oscillations decrease with
magnetic field. Above $\approx 200\,G$, period doubling occurs and
other oscillations are clearly visible. These oscillations have a
period of $100\,G$ corresponding to $\phi_{0}$ per unit cell, and an
amplitude of $\Delta R=2\cdot 10^{-4}\Omega$, corresponding to $9\cdot
10^{-3}\,e^{2}/h$ in terms of dimensionless conductance.

\begin{figure}
\rule{5cm}{0.2mm}\hfill\rule{5cm}{0.2mm}
\vskip 0.5cm
\begin{center}
\psfig{figure=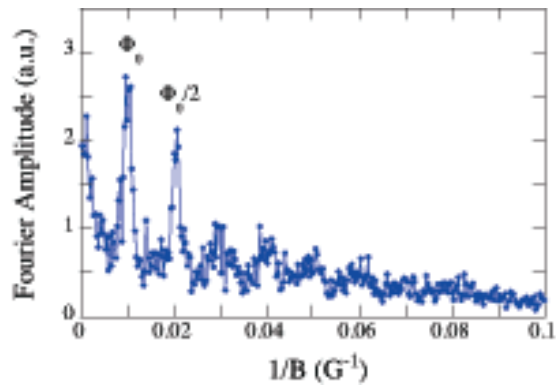,height=5cm}
\end{center}
\rule{5cm}{0.2mm}\hfill\rule{5cm}{0.2mm}
\caption{Fourier spectrum of the magnetoresistance of a metallic 
$\mathcal{T}_{3}$ lattice. The magnetic field range is $0.2-8\,T$.} 
\label{fig:fft}
\end{figure}

Additional magnetoresistance measurements have been performed for
magnetic field up to $8\,T$. A typical Fourier spectrum of this
measurement is diplayed in figure \ref{fig:fft}. In this spectrum,
one clearly observes two peaks at $0.01\,G^{-1}$ and $0.02\,G^{-1}$,
corresponding to $\phi_{0}/2$ and $\phi_{0}$ per unit cell
respectively. At higher frequencies, two or three additional peaks
are slightly visible. It should be stressed that in this spectrum,
the amplitude of the peak at $\phi_{0}$ frequency is larger than that
of the peak at $\phi_{0}/2$ frequency.

Two effects can be invoked in order to explain the observation of
these unexpected $h/e$ oscillations. First, they can be attributed to
reminiscence of the Aharonov-Bohm conductance oscillations in this
(otherwise peculiar) network of rings. If such oscillations have been
observed for a \emph{line} of rings\cite{Washburn}, they have never
been observed in a \emph{network} of rings, due to the lack of
sensitivity and the very large number of rings\cite{Pannetier}. In
this case, the relevant parameter is the ratio between the amplitude
of the $h/e$ (AB) and the $h/2e$ (AAS) component. However, $h/e$
oscillations should vanish like $1/\sqrt{N}$ with $N$ the number of
rings: in our case, both components have roughly the same amplitude. 
Only a detailed calculation of this effect for this particular
geometry should allow meaningful comparison between experiment and
theory\cite{Texier}.

The persistence of $h/e$ oscillations in this $\mathcal{T}_{3}$
lattice could also be attributed to the localisation of the electrons. 
It has been shown theoretically\cite{Vidal2} and proven
experimentally\cite{Naud} that in such a geometry, Aharonov-Bohm cages
lead to a striking robustness of the $h/e$ component of the
conductance oscillations against disorder averaging, at least in the
single channel limit. In this context, it is tempting to consider our
observation as a signature of the Aharonov-Bohm cage effect in the
multi channel limit. The rich content in harmonics observed in the
Fourier spectrum (see figure \ref{fig:fft}) could be another evidence
of this Aharonov-Bohm cage effect. Again, a detailed calculation of
these subtle localisation phenomena in the multi-channel limit is
needed to give a definitive conclusion on our observation. In
addition, measurements on square lattices are necessary for a complete
understanding the observed $h/e$ oscillations.

\section{Conclusion}\label{sec:plac}

We have measured the magnetoconductance of a metallic (gold)
$\mathcal{T}_{3}$ lattice made of $4500$ unitary cells. Superimposed
on the usual Aronov-Altshuler-Spivak oscillations with period $h/2e$,
we observe clear oscillations with period $h/e$ which persist at high
magnetic field up to $8\,T$. The Fourier spectrum of the
magnetoconductance reveals a rich content in harmonics.

This first observation of $h/e$ conductance oscillations in a network
can be attributed either to the reminiscence of the Aharonov-Bohm
oscillations in each cell, or to the geometrical localisation effect
related to the presence of Aharonov-Bohm cages. In both cases, a
detailed theoretical calculation is needed to give a definitive
statement. Additional measurements on networks of different sizes and
topologies ($\mathcal{T}_{3}$ lattices \textit{vs} square lattices)
are necessary to get a deeper comprehension of these subtle
interference effects.

\section*{Acknowledgments}

We thank G. Montambaux, C. Texier, J. Vidal, B. Dou\c{c}ot and L. P.
L\'{e}vy for fruitfull discussions and comments. We acknowledge
financial support the french ministry of science, grant numbers 
NN/0220112 and 0220222, and by the ``\textsc{ipmc}'', Grenoble.


\end{document}